\documentclass{emulateapj}
\shorttitle{CB Disk Scenario for the Negative Orbital-Period Derivative of 4U1820}
\shortauthors{Jiang, Chen \& Li}

\begin{document}


\title{A Circumbinary Disk Scenario for the Negative Orbital-Period Derivative of the
Ultracompact X-ray Binary 4U1820$-$303}


\author{Long Jiang$^{1,2}$, Wen-Cong Chen$^{1,3,4}$, Xiang-Dong Li$^{2,5}$}

\affil{$^{1}$ School of Physics and Electrical Information, Shangqiu Normal University, Shangqiu 476000, China;\\
$^{2}$ Key laboratory of Modern Astronomy and Astrophysics (Nanjing University), Ministry of Education, Nanjing 210046, China; \\
$^{3}$ Department of Physics, University of Oxford, Oxford OX1 3RH, UK; chenwc@pku.edu.cn\\
$^{4}$ Argelander-Insitut f\"{u}r Astronomie, Universit\"{a}t Bonn, Auf dem H\"{u}gel 71, 53121 Bonn, Germany;\\
$^{5}$ Department of Astronomy, Nanjing University, Nanjing 210046, China; lixd@nju.edu.cn}



\begin{abstract}
It is generally thought that an ultracompact X-ray binary is composed
with a neutron star and a helium white dwarf donor star.
As one of the most compact binaries, 4U 1820-303 in globular cluster NGC 6624 was predicted
an orbital-period derivative at a rate of $\dot{P}/P\sim1.1\times10^{-7}$ ${\rm yr^{-1}}$
if the mass transfer is fully driven by gravitational radiation.
However, the recent analysis of the 16 ${\rm yr}$ data
from \textit{Rossi X-ray Timing Explorer}
and other historical records yielded a negative orbital-period derivative in the past $35$ yr.
In this work, we propose an evolutionary circumbinary (CB) disk model to account for this anomalous orbital-period derivative.
4U 1820-30 was known to undergo superbust events caused by runaway thermal nuclear burning on the neutron star. We assume that for a small fraction of the superbursts, part of the ejected material may form a CB disk around the binary. If the recurrence time of such superbursts is $\sim10,000$ yr and $\sim10$\% of the ejected mass feeds a CB disk, the abrupt angular-momentum loss causes a temporary orbital shrink, and the donor's radius and its Roche-lobe radius do not keep in step.
Driven by mass transfer and angular-momentum loss, the binary would adjust its orbital parameters to recover a new stable stage.
Based on the theoretical analysis and numerical simulation, we find that
the required feed mass at the CB disk is approximately $\sim 10^{-8}$ ${\rm M_{\odot}}$.
\end{abstract}


\section{Introduction}
4U 1820$-$303 (hereafter 4U 1820) located in the globular cluster NGC 6624 is an ultracompact X-ray binary (UCXB),
which is defined by an ultra-short orbital period (usually less than 1 hour).
Since its first discovery as a bright X-ray source \citep[]{gia1974},
4U 1820 has been extensively observed with many X-ray telescopes.
Its peak-to-peak modulation period $\sim 685$ {\rm s} \citep[]{stel1987, sans1989, ande1997}
is generally believed as the orbital period of a white dwarf orbiting a neutron star \citep[]{rapp1987}.

As the most compact UCXB, the formation and evolution history of 4U 1820 has been widely studied so far.
Considering the high stellar density of globular cluster,
\cite{verb1987} suggested that this system was formed via spiral-in phase of
a neutron star into a red giant after their direct collision.
Taking common envelope spiral-in into consideration, \cite{bail1987}
calculated an evolutionary sequence from 4U 2127$+$12 in M15 to 4U 1820.
In their scenario, the binary system was produced via tidal capture
between a main-sequence star and a neutron star.
Similarly, \cite{rasio2000} proposed that this source originated from an exchange interaction between a
neutron star and a primordial binary including two main sequence stars,
in which a common envelope evolution phase would subsequently be expected.

Although adopting two different formation channels for 4U 1820
\cite{verb1987} and \cite{bail1987} agreed
that the mass transfer is dominantly driven by gravitational radiation.
Based on a similar model, \cite{rapp1987} calculated the evolutionary sequence of 4U 1820,
and found that its donor star is a helium white dwarf of mass $\sim0.06-0.08$ ${\rm M}_\odot$
with a radius within $\sim20\%$ of a completely degenerate configuration.
They also predicted an X-ray luminosity of $L_{\rm X}\sim8\times10^{37}$ ${\rm erg}$ ${\rm s}^{-1}$,
and an orbital-period derivation of $\dot{P}/P \sim 1.1 \times 10^{-7}$ ${\rm yr}^{-1}$
driven by mass transfer from the white dwarf to the neutron star.

However, \cite{sans1989} obtained an orbital-period derivative
as $\dot{P}/P \sim-6\times 10^{-8}$ ${\rm yr}^{-1}$ by Ginga observations.
Subsequently, \cite{Tan1991} also reported a negative period derivative at a significance of $99.9\%$.
Employing simultaneous ROSAT/Ginga observations, \cite{Klis1993a}
reported an average orbital-period derivative $\dot{P}/P \sim(-0.88 \pm 0.16)\times 10^{-7}$ ${\rm yr}^{-1}$.
Based on the further observation by ROSAT, \cite{Klis1993b}
obtained $\dot{P}/P= (-5.3 \pm 1.1) \times 10^{-8}$ ${\rm yr}^{-1}$.
\cite{chou2001} also reported a negative derivative of $\dot{P}/P\sim-3.5\times10^{-8}$ ${\rm yr^{-1}}$.
Recently, \cite{peut2014} analyzed the 16 yr \textit{Rossi X-ray Timing Explorer}
(\textsl{RXTE}) data of 4U 1820, and refined the negative orbital-period derivative
to be $\dot{P}/P= (-5.3 \pm 0.3) \times 10^{-8}$ ${\rm yr}^{-1}$ at a $>17\sigma$ level.
Obviously, all of the observations confirmed a negative orbital-period derivative for 4U 1820,
which is contrary to the theoretical prediction given by \cite{rapp1987}.

How to explain the difference in the orbital evolution between the binary evolutionary theories and the observations?
To resolve this problem, many authors proposed various scenarios, which can be divided into two categories.
In the first one it is thought that the orbital-period derivative arises from the binary evolution.
Considering the secondary is a non-degenerate helium-star of $0.6$ ${\rm M_\odot}$
orbiting a $1.3$ ${\rm M_\odot}$ neutron star, \cite{savo1986}
calculated the evolution of a compact binary with initial period 37 minutes.
They found that the orbital period could be as low as 11 minutes
and a negative period derivative could be attained also.
Their calculation indicates that the donor star is a helium-burning star
with mass of $\sim0.24$ ${\rm M_\odot}$, which is not expected to exist
in old stellar population like globular clusters \citep{stel1987, verb1987}.
Assuming that stars more massive than the turn-off mass could be
formed during the close encounter in globular clusters,
\cite{Klis1993a} restudied the possibility of Helium-burning donor star,
and obtain a conclusion similar to \cite{savo1986}.
\cite{Ma2009} found that a circumbinary disk (CB disk) around the binary can drive low-mass X-ray binaries to
ultra-short periods as short as 6 min.
Recently, \cite{chen2016} investigated an alternative formation channel toward UCXBs
with an orbital period of 11 minutes, which evolved from intermediate-mass X-ray binaries
driven by magnetic braking of Ap/Bp stars.
Their simulations also indicate a long-term period-decreasing phase.

In the other case, some researchers suggested that the apparent orbital period derivative can
be interpreted by an accelerating motion of the binary towards observers.
\cite{Tan1991} argued the acceleration may
arise from a third body or the cluster potential.
\cite{peut2014} explored the possibility of a stellar mass remnant
close to 4U 1820 or an intermediate-mass black-hole inside NGC 6624.
\cite{Klis1993b} even suggested that the secular period change was dominated by
systematic or random change of position and shape of occulting bulge on the accretion disk rim.

Taking the orbital period change as the result of binary evolution,
here we suggest that the negative orbital period derivative of 4U 1820 could be produced by a temporary CB disk,
which can extract the orbital angular momentum via resonant interaction from the binary system.
Our theoretical analysis and numerical simulation indicate that
the CB disk mass that can account for the observed orbital-period derivative is approximately
$\sim1.0-1.5\times10^{-8}$ ${\rm M}_{\odot}$.

In the following section, we present a simple theoretical analysis for the current orbital-period derivative of 4U 1820,
and describe the CB disk model in section 3.
The numerical simulation method and simulated results are described in detail in section 4. In section 5, we present a brief summary and discussion.

\section{Theoretical analyse of the current orbital period derivative of 4U 1820}

Considering a binary consisting of a neutron star (of mass $M_1$) and
a He white dwarf secondary (of mass $M_2$) in a circular orbit,
the orbital angular momentum is $J=2\pi\mu a^2/P$.
Here $\mu=M_1M_2/(M_1+M_2)$ is the reduced mass, and $a$ is the orbital separation.
The spin angular momentum of the donor star is neglected because
it is much less than the orbital angular momentum.
By a simple logarithmic differentiation of the
angular momentum, we have
\begin{equation}
\frac{\dot{P}}{P}=3\frac{\dot{J}}{J}+\frac{\dot{M}_1+\dot{M}_2}{M_1+M_2}-3\frac{\dot{M}_1}{M_1}-3\frac{\dot{M}_2}{M_2}.
\end{equation}

Taking the Eddington accretion rate $\dot{M}_{\rm Edd}$ into account,
the accretion rate of the neutron star is assumed to be $\dot{M}_1={\rm min}(f_1|\dot{M}_2|, \dot{M}_{\rm Edd})$, where $f_1\leq1$ is a constant. The accretion efficiency of the neutron star is defined as
\begin{equation}
\beta\equiv\dot{M}_1/|\dot{M}_2|.
\end{equation}
Therefore, when $\dot{M}_{\rm Edd}<f_1|\dot{M}_2|$, $\beta<f_1$ while $\dot{M}_{\rm Edd}\geq f_1|\dot{M}_2|$, $\beta=f_1$. X-ray observation performed by \cite{stel1987} reported an X-ray luminosity of $2-10\times10^{37}$ ${\rm erg}$ $\rm s^{-1}$, which implies the accretion rate of 4U 1820 $\dot{M}_1\sim10^{-8}$ ${\rm M}_{\odot}$ $\rm yr^{-1}$.

Magnetic braking would turn off for a fully convective star \citep{rapp83,spru83},
hence we consider angular momentum loss due to gravitational radiation and mass loss:
\begin{equation}
\frac{\dot{J}}{J}=\frac{\dot{J}_{\rm GR}}{J}+\frac{\dot{J}_{\rm ML}}{J},
\end{equation}
where $\dot{J}_{\rm GR}$, and $\dot{J}_{\rm ML}$ are the angular-momentum-loss rate by gravitational radiation, and mass loss, respectively. The angular-momentum-loss rate by gravitational radiation is given by \cite{land1975}:
\begin{equation}
\frac{\dot{J}_{\rm GR}}{J}=-\frac{32(2\pi)^{8/3}}
{5c^5}G^{5/3}M_1M_2M_{\rm T}^{-1/3}P^{-8/3},
\end{equation}
where $c$ is the light velocity, $G$ is the gravitational constant, and
$M_{\rm T}=M_{\rm 1}+M_{\rm 2}$ is the total mass of the binary.
Taking $M_1=1.58$ ${\rm M}_{\odot}$ \citep{guve2010}, $M_2=0.07$ ${\rm M}_{\odot}$  \citep{rapp1987}
and $P_{\rm orb}=685$ ${\rm s}$ \citep[]{stel1987, sans1989, ande1997},
we can derive ${\dot{J}_{\rm GR}}/{J}\sim-10^{-7}$ $\rm yr^{-1}$.

The angular-momentum-loss rate by mass loss can be written as
\begin{equation}
\dot{J}_{\rm ML}=2\pi j(1-\beta)\dot{M}_2a^2/P,
\end{equation}
where $j$ is the specific angular momentum of the ejected matter in units of $2\pi a^2/P$.
In this work, the mass loss during the accretion is assumed to form an isotropic wind in the vicinity of the neutron star, and carry away its specific orbital-angular momentum, i.e. $j=M^2_{\rm 2}/(M_{\rm 1}+M_{\rm 2})^2$.
Based on some parameters mentioned above, one can find ${\dot{J}_{\rm ML}}/{J}\sim-10^{-11}$ $\rm yr^{-1}$.
Comparing with gravitational radiation, the angular-momentum loss due to mass loss can be ignored.

Based on some parameters mentioned above, Equation (1) can yield the current orbital period derivative of 4U 1820.
For a conservative mass transfer, the current period derivative
should be $\dot{P}/P\sim10^{-7}$ ${\rm yr^{-1}}$, while $\dot{P}/P\sim5\times10^{-7}$ ${\rm yr^{-1}}$
for nonconservative mass transfer with $f_1=0.5$ \citep{posi02}.
Similar to the discussion given by \cite{rapp1987},
both cases predict a positive orbital period derivative, which is contradicted with the observation.

\section{A CB disk model}

To interpret the negative orbital-period derivative observed in 4U 1820, here we propose an evolutionary CB disk model.
The influence of CB disk on the evolution of cataclysmic variables \citep{spru01,taam01}, black-hole X-ray binaries \citep{chen2006b,chen2015}, Algol binaries \citep{chen2006a}, and UCXBs \citep{Ma2009}
have been studied extensively.
All these works show that CB disk can efficiently extract the orbital angular momentum from the binary system, and enhance the mass transfer rate and accelerate the evolution process.
In this work, we adopt a different CB disk model.
The resonant interaction between the binary and the CB disk has been well studied
\citep[]{arty1991, lubow1996, lubow2000, derm2013},
and its predicted orbital angular momentum loss rate is given by \citep{lubow1996}:
\begin{equation}
\frac{\dot{J}_{\rm CB}}{J}=-\frac{l}{m}
\frac{M_{\rm CB}}{\mu}\alpha\left(\frac{H}{R}\right)^2\frac{a}{R}\frac{2\pi}{P}.
\end{equation}
Here $M_{\rm CB}$ is the mass of CB disk, $H$ and $R$ ($=\sqrt{r_{\rm in} r_{\rm out}}$, where $r_{\rm in}$, and $r_{\rm out}$ are the inner and outer radius of the CB disk, respectively) are the thickness
and the half angular momentum radius of the disk, respectively.
$l$ and $m$ are integers describing the binary potential, $\alpha~(=0.1)$ is the viscosity parameter of the disk.

\cite{mira2015} studied the tidal truncation between circumstellar and CB disks in binaries, and
found the inner radius of CB disc is $1.5a<r_{\rm in}<3.5a$.
Following their study, assuming the disk extends from $r_{\rm in}=2.5a$ to $r_{\rm out}=10a$,
the half angular momentum radius of the disk is $R=5a$.
According to the study of \cite{dull2001},
the relative thickness of CB disk near the inner edge is $H/R = 0.1-0.25$.
Assuming the disk is thin and the non-axisymmetric potential perturbations are small
we take $(H/R)^2=1/30$ and $l=m=1$ in this work.

According to above equations a CB disk with mass of $\sim10^{-8}$ ${\rm M}_{\odot}$
can reproduce the observed negative orbital-period derivative
for both conservative and the nonconservative mass transfer cases.
Certainly, the CB disk would slowly induce a small eccentricity \citep{derm2013}.
However, the tidal interaction between the donor star with a relatively deep convective envelope
and the neutron star in such a compact orbit would rapidly circularize the orbit.

We propose the material of the CB disk comes from some \emph{rare} superburst events.
Superbursts are hour-long thermonuclear runaway burning processes which were
first observed by \cite{corn2000} from the neutron star low mass X-ray binary 4U 1735$-$444.
A superburst of 4U 1820 was discovered by \cite{stro2000} and \cite{stro2002}
as the second superburst event ever observed.
Subsequently, many other superburst events have been discovered from other LMXBs
such as KS 1731$-$260, Serpens X$-$1 GX 3$+$1 and so on.
At present, about one dozen superburst sources (candidates)
have been reported, and one third of these sources were observed recurrently \citep[]{keek2012}.
Interestingly, although the recurrence times estimated from observation of other sources are on the order of a year \citep{kuul2004},
the second superburst of 4U1820 was observed to be 11 years after the first event \citep[]{keek2012},
which is consistent to the prediction ($\sim13~\text{yr}$) by \citet{stro2002}.

Compared with more frequent type I bursts with short duration (typically $\sim20$ s),
superbursts release X-ray energy $\sim10^{42}$ ${\rm erg}$ during an event,
which is three orders of magnitude higher than that of type I burst.
Considering a substantial fraction of the released energy during the event
is carried away by neutrinos, and is conducted to the inner part of the neutron star,
the total energy released during the event is much larger $>10^{43}$ ${\rm erg}$ \citep{stro2002}.
Observational and theoretical studies
indicate that it is a thermonuclear runaway burning process,
and the energy source is most likely carbon and/or oxygen on the surface of the neutron star \citep{stro2002,zand2012}.
A detailed analysis indicated the mass of the burning-carbon layer
is probably $\geq10^{26}$ ${\rm g}$ \citep{stro2002}.

In principle, there exist a range for the released energy of different superburst events \citep{kuul2002}.
Here we assume that some peculiar superburst events (with a probability of $0.1\%$, hereafter rare superbursts) in a layer of $\sim 10^{27}$ ${\rm g}$ ($5\times10^{-7}$ ${\rm M}_{\odot}$) may be responsible for the formation and evolution of CB disk.
In our calculation, 20 percent of the burning-carbon layer was assumed to escape from the neutron-star surface,
and 10 percent of the ejected material (2 percent of total burning mass,
i.e. $\bigtriangleup M_{\rm CB}=1.0\times10^{-8}$ ${\rm M}_{\odot}$)
suddenly feed the CB disk \footnote{The viscous timescale of a typical CB disk is of order of years, much shorter than the evolutionary timescale considered here.} during every rare-superburst event. Take the recurrence time of superburst events of 4U1820 is 10 yr, as a result, the CB disk would experience a mass feed with $\bigtriangleup M_{\rm CB}$ every $\sim10,000~\text{yr}$. Meanwhile, the mass of CB disk is assumed to decrease at a rate $0.1\%M_{\rm CB}~\rm yr^{-1}$ (the CB disk only remains $\sim5\times10^{-5}\bigtriangleup M_{\rm CB}$ at the beginning of next rare superburst) due to the photo-evaporation of the neutron star's spin-down energy \citep{anto14}.
According to Equation (5), the angular-momentum loss due to sudden mass loss from the neutron-star surface induced by (normal or rare) superburst is $\sim-(10^{-10}-10^{-11})J$ $\rm yr^{-1}$, and can be ignorable.

\section{Numerical simulation}

To evaluate the CB disk scenario in detail, we develop a fast binary-evolution program for 4U 1820.
As shown in Fig. 1, our calculation starts from gravitational wave emission dominated detached binary consisting of a neutron star and a low-mass He white dwarf.
To fit the current mass of the neutron star ($1.58$ ${\rm M}_{\odot}$, \cite{guve2010}) and the donor star (
$0.07$ ${\rm M}_{\odot}$, \cite{rapp1987}),
the initial input parameters is set to $M_1=1.5$ ${\rm M}_{\odot}$,
$M_{2}=0.15$ ${\rm M_\odot}$, and orbital period $P=0.2$ hr.
Such an orbital-period can ensure a He white dwarf to decouple from its Roche lobe and
the steady mass transfer condition is not introduced manually but satisfied via slowly adjust of gravitational wave emission.
This is different from \cite{rapp1987} where the evolution was studied from Roche lobe filled binary with simple relation between the mass of white dwarf donor and the orbital period: $P\varpropto M_2^{-1}$.

\begin{figure}
\centering
\includegraphics[width=0.5\linewidth,trim={30 0 80 0}]{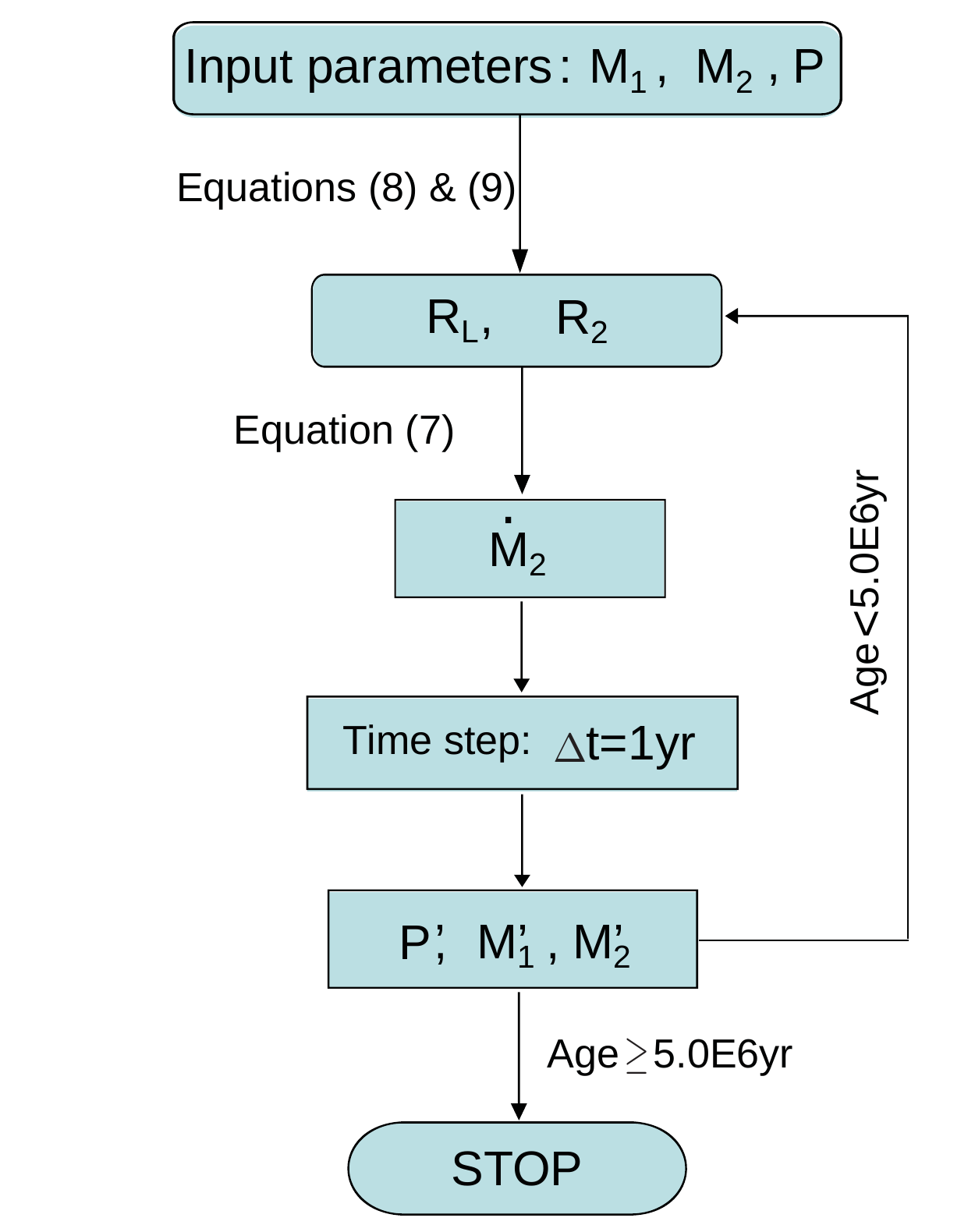}
\caption{\label{fig:per-m2} Simulation schedule of 4U 1820.
Parameters and equations used in the simulation see the text for detail.
Parameters without mentioned in the text have the usual meanings.}
\end{figure}

Once the donor star overflows its Roche lobe due to gravitation radiation,
it would transfer material to the neutron star at a rate \citep{nels2001}:
\begin{equation}
\dot{M}_{\rm 2}=-f_2{\rm log}^3\left(\frac{R_2}{R_{\rm L}}\right),
\end{equation}
where $f_2$ is a constant factor.
In calculation, we take $f_2=5\times10^{-3}\text{M}_{\odot}\text{yr}^{-1}$.
$R_2$ is the radius of the secondary, and $R_{\rm L}=af(q)$ is the Roche-lobe radius of the donor star.
The function $f(q)$ only relate to the mass ratio $q=M_2/M_1$.
In this paper we adopt the approximately description given by \citet{egg83}, i.e.
\begin{equation}
R_L=\frac{0.49q^{2/3}}{0.6q^{2/3}+ {\rm ln}(1+q^{1/3})}a.
\end{equation}
Utilizing a simple polytropic index $n=3/2$, the donor-star radius is given by \citep{chan1939}
\begin{equation}
R_2=0.0128(1+X)^{5/3}f_3\left(\frac{M_2}{\rm M_\odot}\right)^{-1/3}{\rm R_\odot},
\end{equation}
where $X$ is the mass abundance of hydrogen ($X=0$ for a pure helium white dwarf in this paper),
and $f_3\geq1$ ($f_3=1$ in this paper) is the radius ratio between the donor star
and the white dwarf with same mass, which is completely degenerate
and only supported  by the Fermi pressure of the electrons.

To test the influence of input parameters on the evolution and compare the results with other works,
we have run four different models as follows:
\begin{itemize}
\item Model 1:$f_1=1.0$, $\bigtriangleup M_{\rm CB}=0$;
\item Model 2:$f_1=1.0$, $\bigtriangleup M_{\rm CB}=1\times10^{-8}$ ${\rm M}_{\odot}$;
\item Model 3:$f_1=0.5$, $\bigtriangleup M_{\rm CB}=0$;
\item Model 4:$f_1=0.5$, $\bigtriangleup M_{\rm CB}=1\times10^{-8}$ ${\rm M}_{\odot}$;
\end{itemize}

\begin{figure}
\centering
\includegraphics[width=0.8\linewidth,trim={30 0 80 0}]{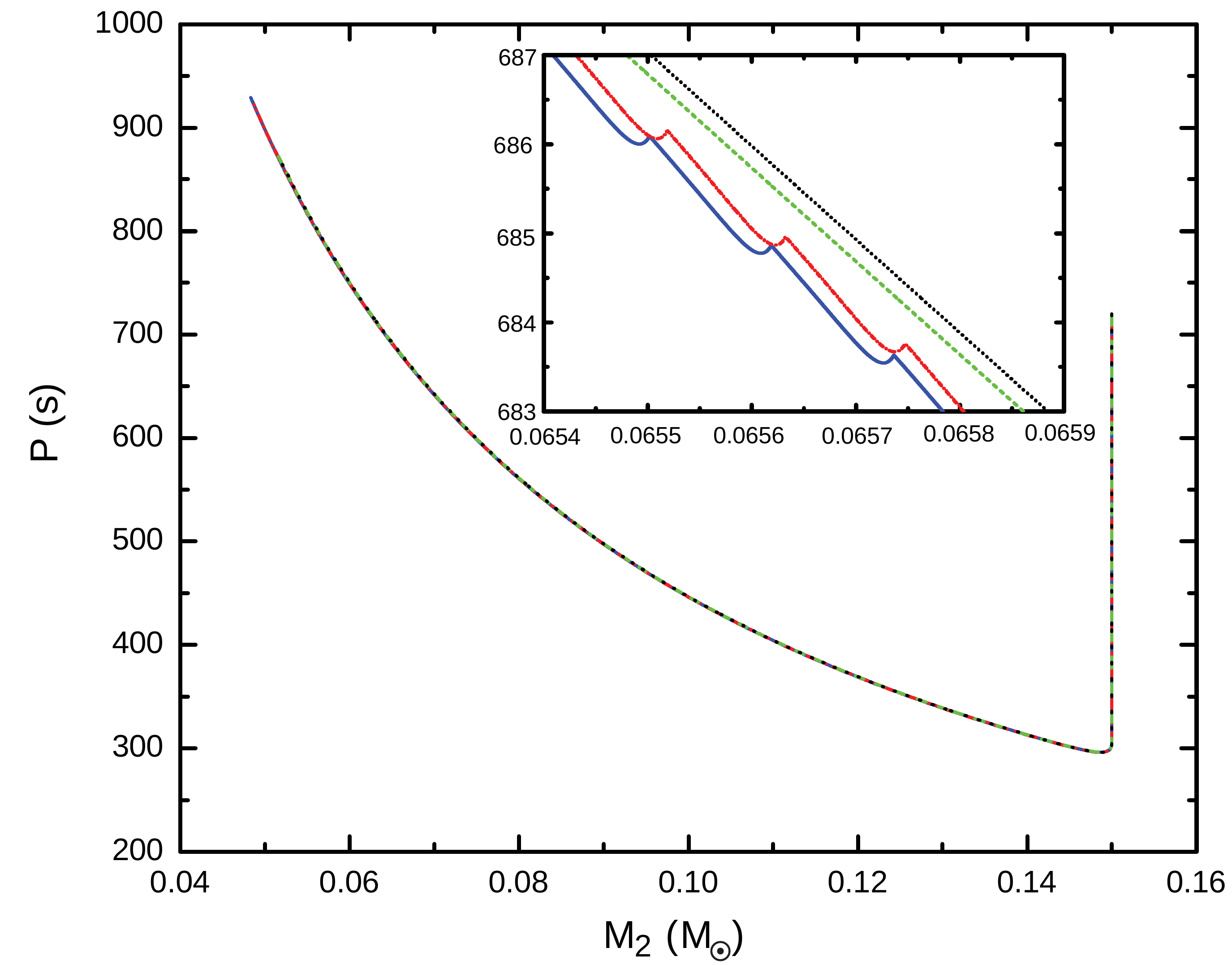}
\caption{\label{fig:per-m2} Orbital period as the function of the donor star
  mass for an UCXB with initial mass $M_{\rm 1}=1.5$ ${\rm M}_{\odot}$,
  $M_{\rm 2}=0.15$ ${\rm M}_{\odot}$, and an initial
  orbital period of 0.2hr. The green dashed, blue solid, black dotted, and red dashed-dotted curves represent the evolution of Models 1, 2, 3, and 4, respectively. }
\end{figure}

Fig. 2 shows the evolution of an UCXB under different models in the $P - M_{2}$ plane.
It is clear that the orbital-period evolution can be divided into two stages.
In the first stage, gravitational radiation dominates the angular momentum loss of the binary,
and causes the orbit to sharply shrink.
After the donor star overflows its Roche lobe, the orbital decay becomes slower
because the material is transferred from the less massive donor star to the more massive accretor.
Subsequently, the orbital period gradually increases, and the second stage starts
where the evolution roughly obeys an orbital period-donor mass relation as $P\varpropto M_2^{-1}$.
Our simulation also shows that at the orbital period of $\sim 11$ min the donor star mass is in the range of $\sim 0.06-0.07~\rm M_{\odot}$, in good agreement with the predication by \cite{rapp1987}.
The CB disk formed by the first rare superburst is assumed to experience a mass feed if the mass-transfer rate $\dot{M}_2>1.0\times10^{-8}~\text{M}_{\odot}\text{yr}^{-1}$.
It seems that four models have a similar evolutionary tracks. However, in the mini panel of Fig. 2, models 2 and 4 predict short-term orbital-decay episodes, which were
induced by sudden mass feed of the CB disk.

\begin{figure*}
\centering
\begin{tabular}{cc}
\includegraphics[width=0.45\textwidth,trim={0 0 0 0}]{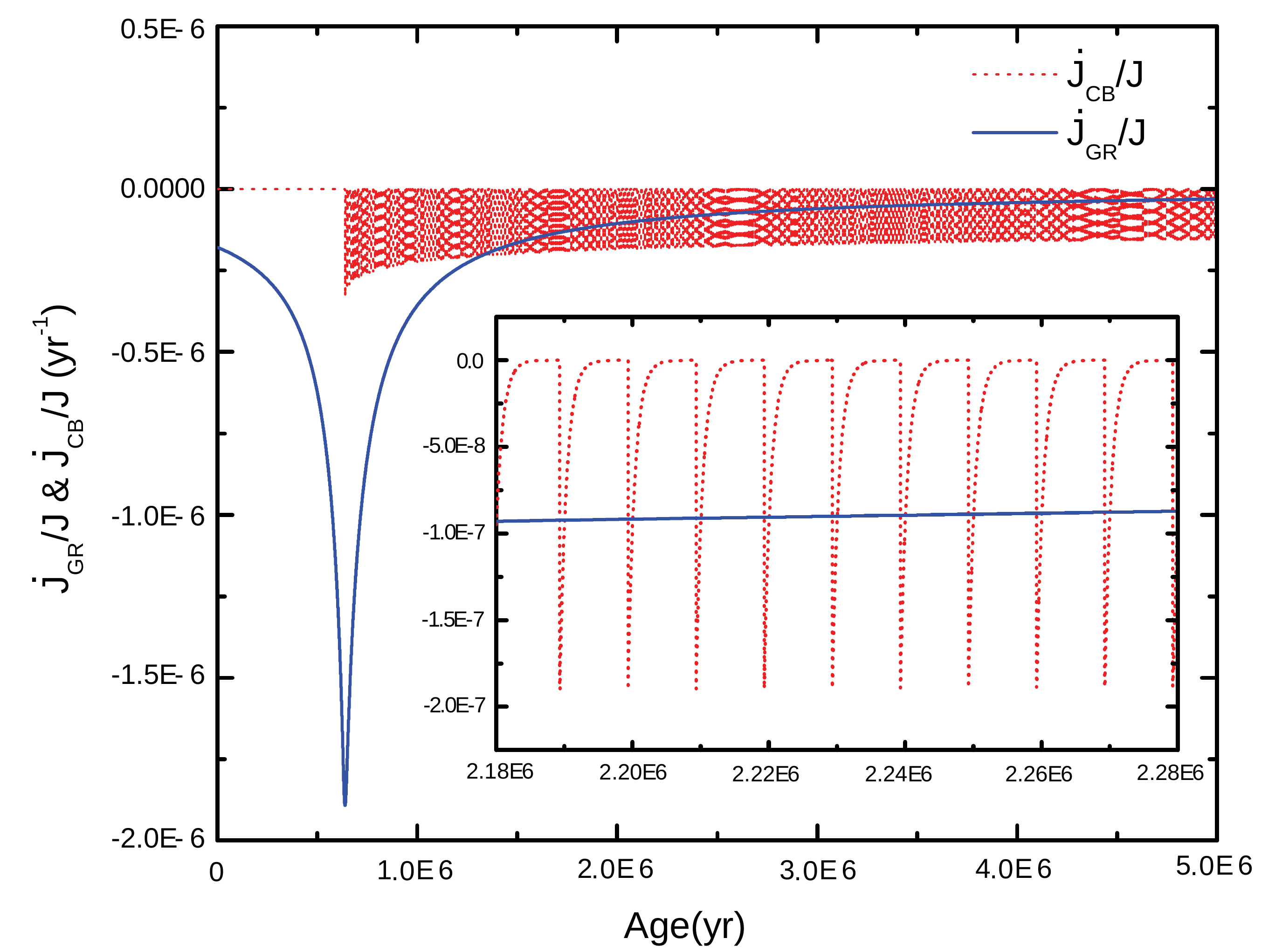} &
\includegraphics[width=0.45\textwidth,trim={0 0 0 0}]{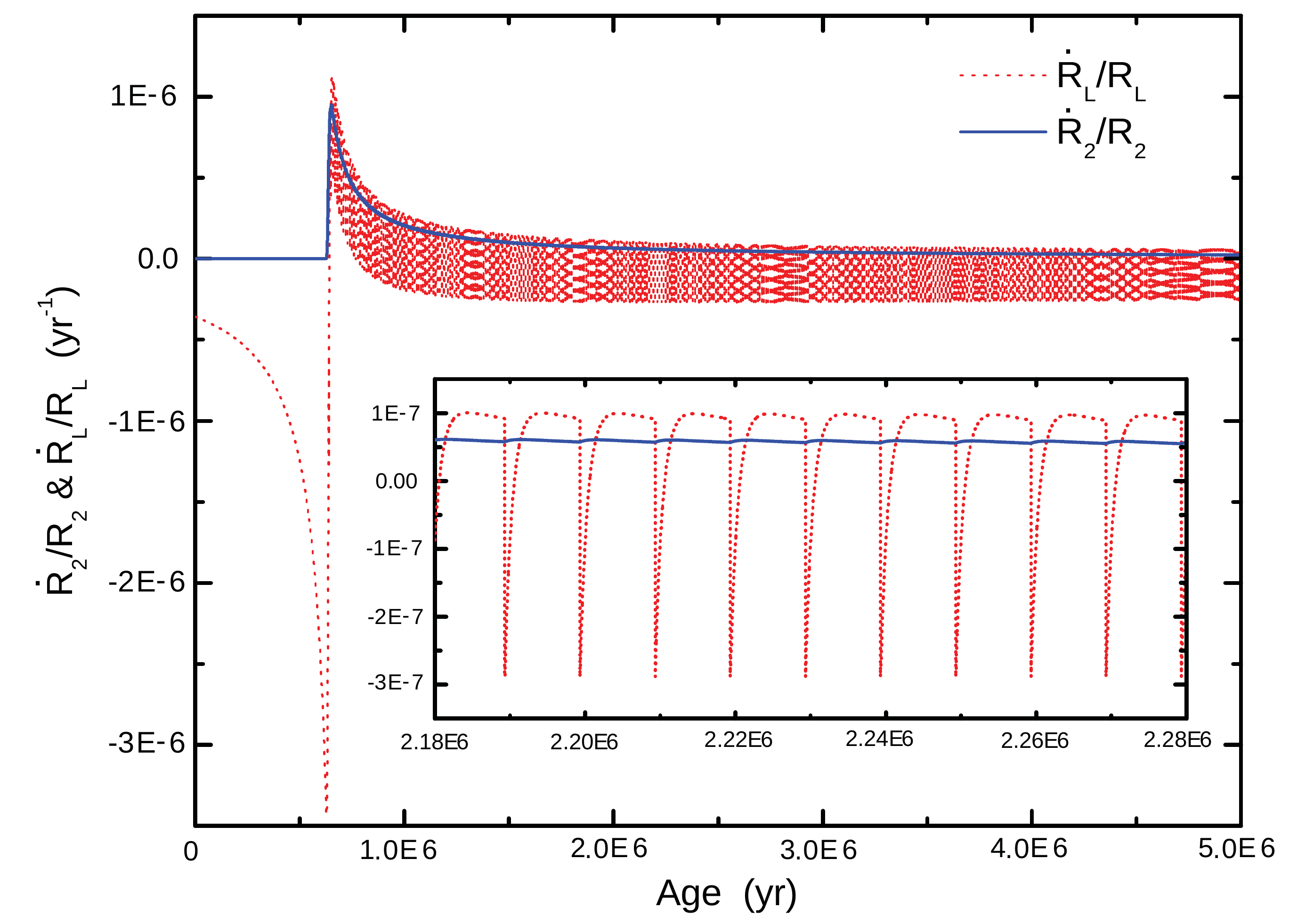} \\
\end{tabular}
\caption{\label{fig:pulses} Evolution of angular-momentum-loss rate (left panel) induced by gravitational radiation and CB disk,  and evolution of derivative rate of donor-star radius and Roche-lobe radius of donor star (right panel) for Model 2.}
\end{figure*}

To interpret the short-term orbital-decay episodes in detail, the evolutions of angular-momentum-loss rate induced by gravitational radiation and CB disk for Model 2 are shown in the left panel of Fig. 3.
Comparing with the continuous evolution of $\dot{J}_{\rm GR}/J$, the sudden mass feedings at CB disk induce sharp dips of $\dot{J}_{\rm CB}/J$.
As shown in the right panel of Fig. 3, for Model 2, the sudden increase of angular-momentum-loss rate induces the sudden decrease of Roche-lobe radius of the donor star. Since the radius of the white dwarf increases gradually due to mass loss,
the stable mass transfer condition $\dot{R}_2/R_2=\dot{R}_{L}/R_{L}$ \citep{Di2008} is ruined.
With the mass decay of CB disk, the angular-momentum-loss rate gradually decreases, and the stable mass-transfer condition reestablished again before next mass feeding.

In Fig. 4, we summarize the evolution of the donor mass, the mass transfer rate, the orbital period,
and the orbital-period derivative.
Some main results can be summarized as follows:

1. Our calculation show that, after each mass feed of rare superburst there exists a short-term orbital-period-decrease phase with 900 yr for Models 2, and 4. Roughly speaking there is a 9\% probability to see 4U1820 in orbital decay.

2. After the recovery of a stable mass transfer, both Models 2 and 4 yield a higher orbital-period derivative than models 1 and 3. This discrepancy origins from an additional mass loss related to the formation of a CB disk.

3. A CB disk can efficiently extract the orbital-angular momentum from the binary, accelerate the evolutionary process, and result in a smaller donor mass and a higher mass-transfer rate.

4. The current mass-transfer rate of donor star is $\sim1.1-1.2\times10^{-8}~\text{M}_{\odot}\text{yr}^{-1}$.
This indicates an X-ray luminosity of $\sim5-12\times10^{37}~\text{erg}\,\text{s}^{-1}$
which is consistent with the observation of $\sim2-10\times10^{37}~\text{erg}\,\text{s}^{-1}$ \citep{stel1987}.
During the recovery of a stable mass transfer, both the mass-loss rate of donor star and the absolute value of orbital-period derivative should decrease.

5. We also simulate the evolution when the feed mass onto the CB disk
$\bigtriangleup M_{\rm CB}=10^{-9}$ ${\rm M}_{\odot}$, and $10^{-7}$ ${\rm M}_{\odot}$.
Compared to the case when $\bigtriangleup M_{\rm CB}=10^{-8}$ ${\rm M}_{\odot}$, the evolution in the former case is similar but there is no orbital-period-decrease phase since the CB disk mass is much lower.
In the latter case, with a heavier CB disk induces a much faster evolution with a higher mass-transfer rate.
At $P\sim685~\text{s}$, the evolutionary age and the mass-transfer rate are $\sim1.2\times10^6~\text{yr}$, and $\sim2\times10^{-8}~\text{M}_{\odot}\text{yr}^{-1}$, respectively. Meanwhile, the simulation produce longer timescales ($\sim1500~\text{yr}$) in the orbital-period-decrease phases, and more abrupt orbital-period-derivative dips ( $\sim1\times10^{-6}~\text{yr}^{-1}<\dot{P}/P<\sim-5\times10^{-6}~\text{yr}^{-1}$).
\begin{figure*}
\centering
\begin{tabular}{cc}
\includegraphics[width=0.45\textwidth,trim={0 0 0 0}]{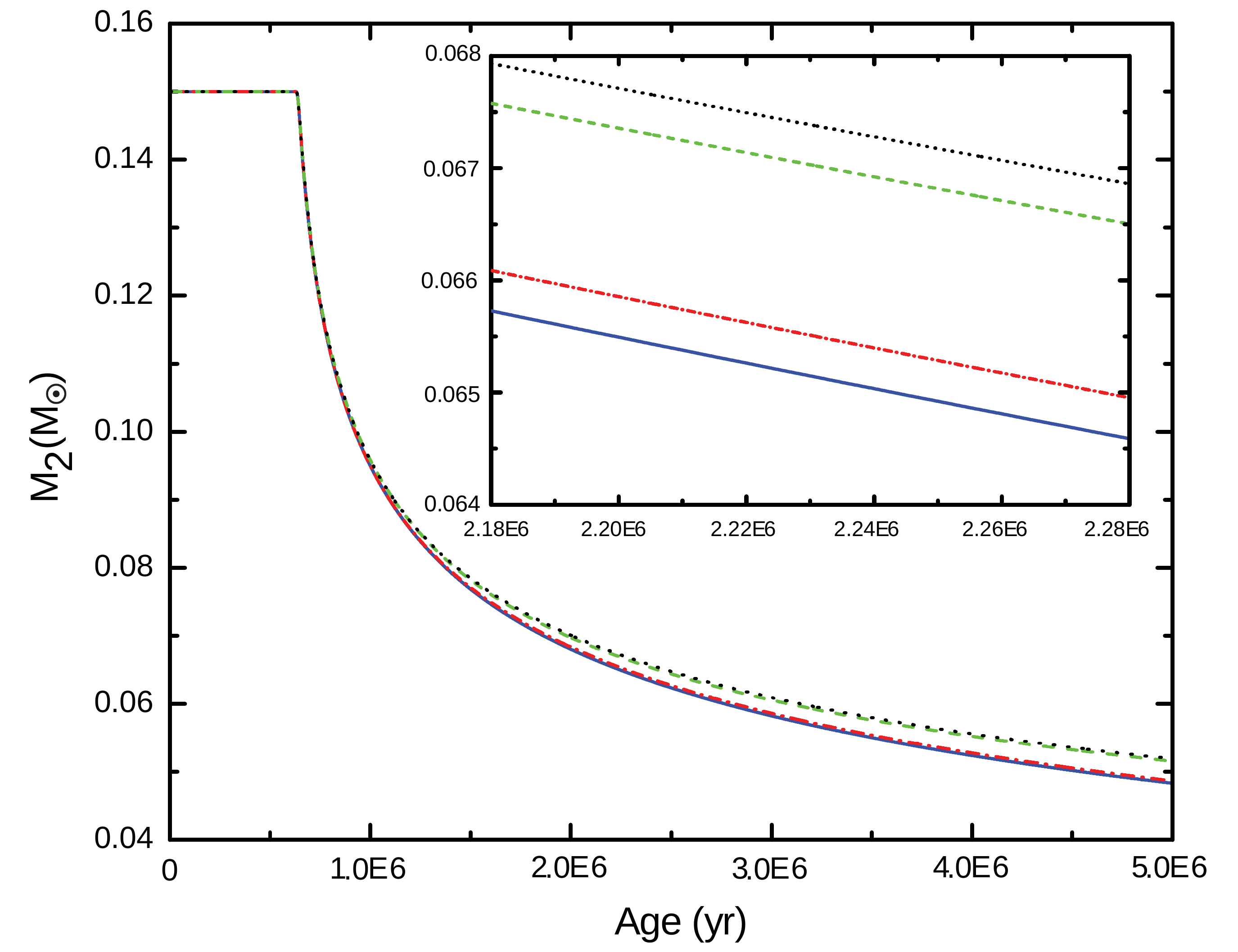} &
\includegraphics[width=0.45\textwidth,trim={0 0 0 0}]{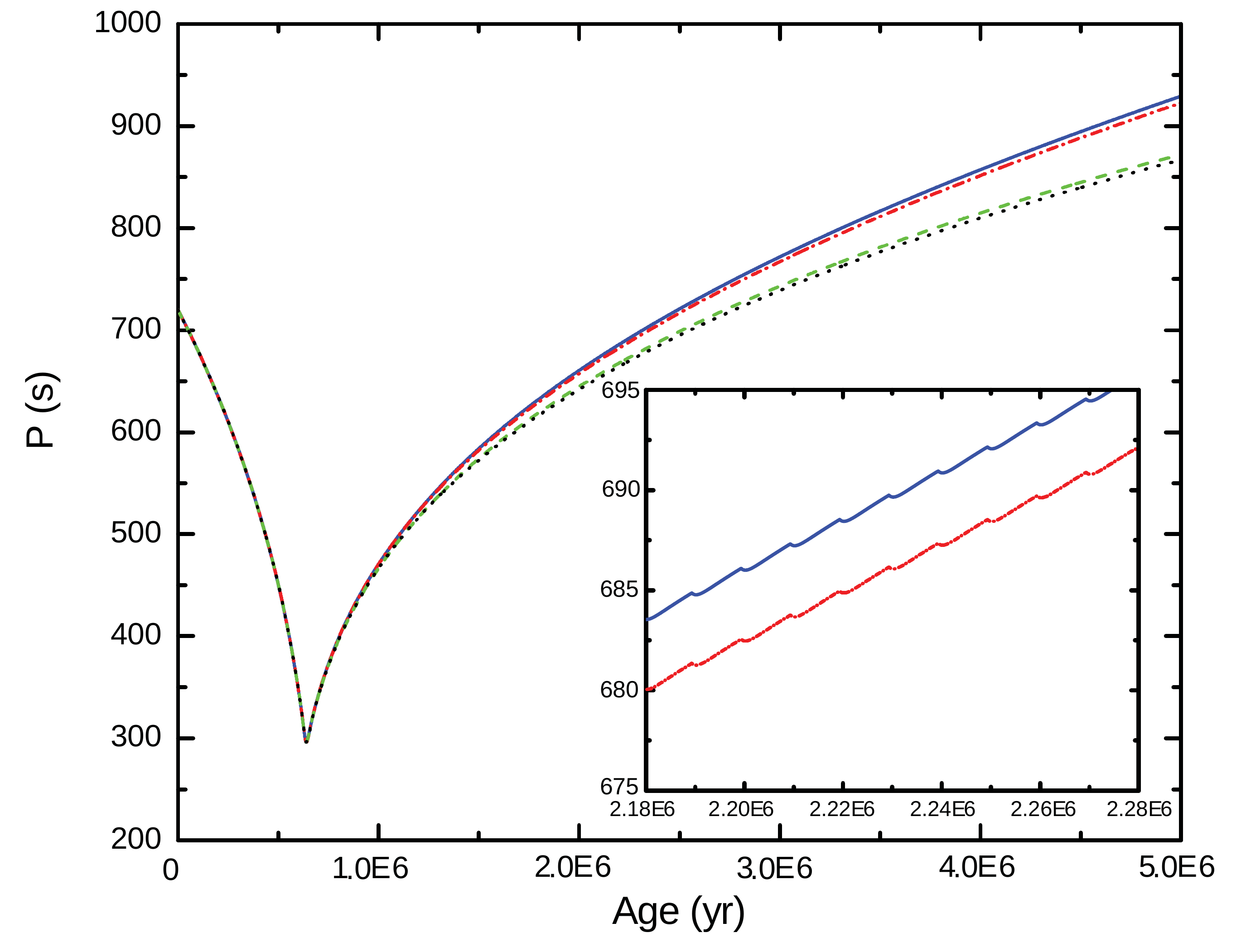} \\
\includegraphics[width=0.45\textwidth,trim={0 0 0 0}]{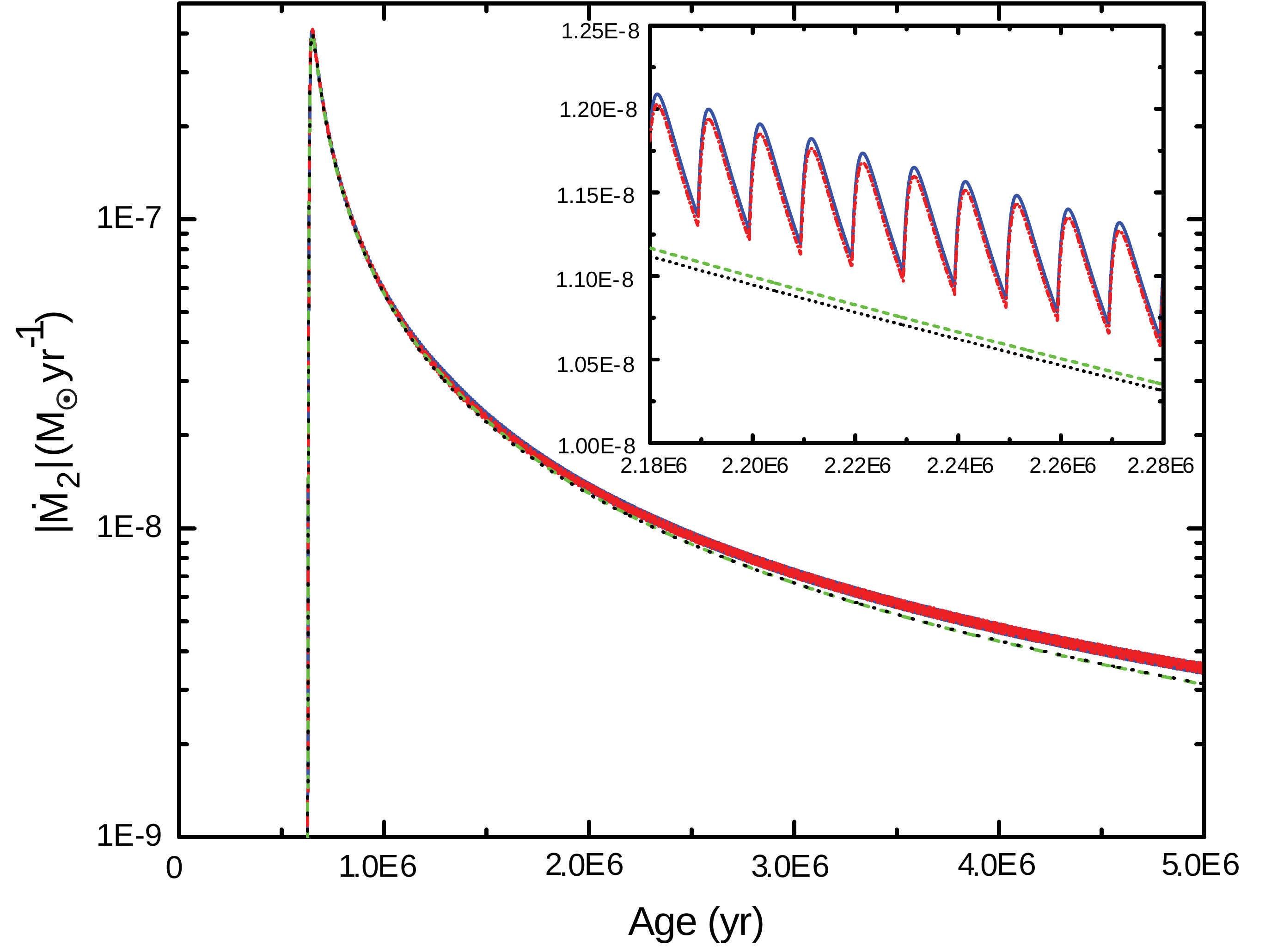} &
\includegraphics[width=0.45\textwidth,trim={0 0 0 0}]{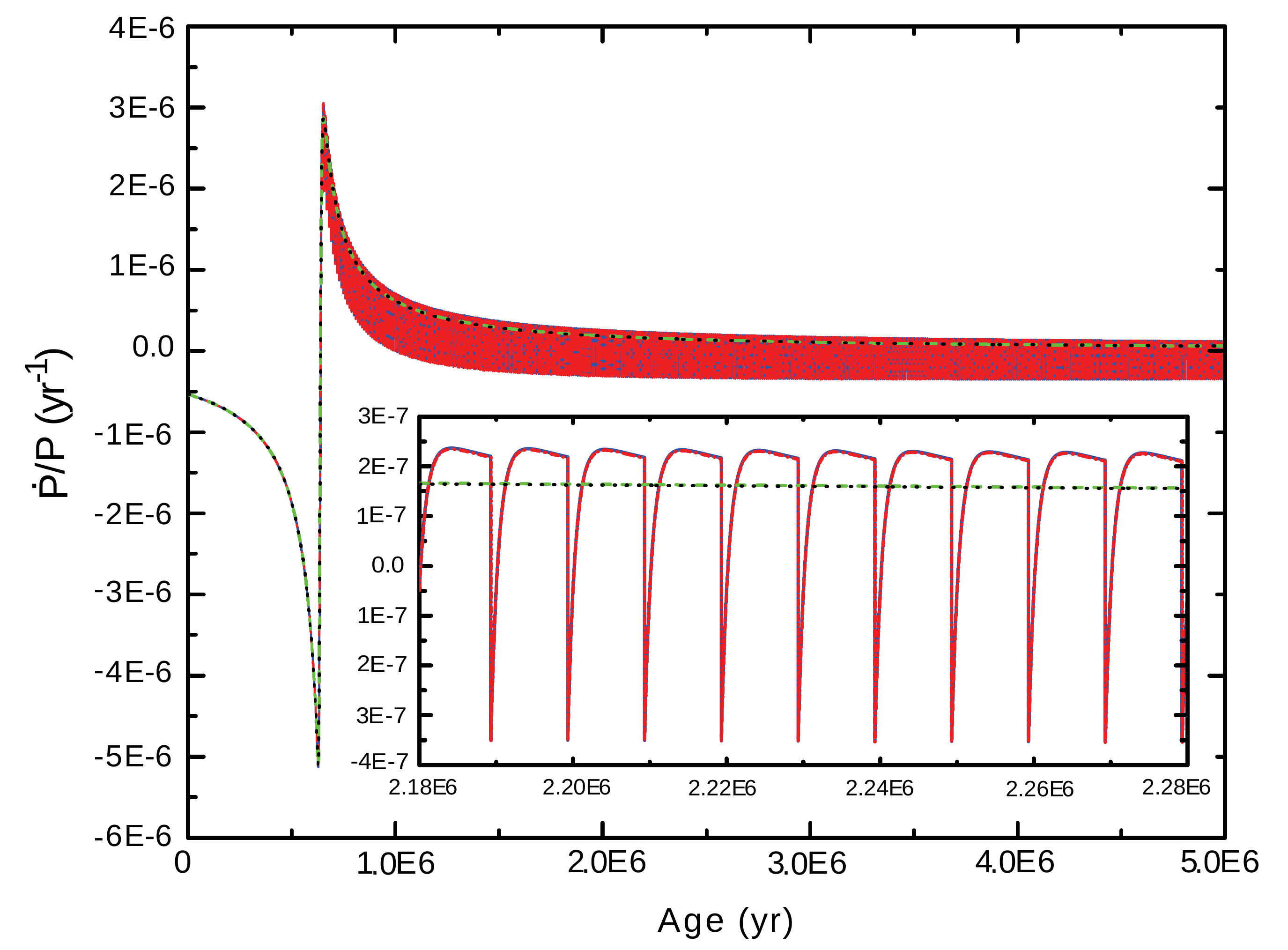} \\
\end{tabular}
\caption{\label{fig:pulses} Evolution of donor mass (top left panel), mass loss rate (middle left panel), orbital period (top right panel), and orbital-period derivative (middle right panel) for an UCXB with a same initial parameters as Fig. 2.
The meaning of different curves also same to Fig. 2.}
\end{figure*}

\section{Summary and Discussion}

Mass transfer from the less massive donor star to the more massive accretor always induces an expanding orbit.
In the case of stable mass transfer, the orbital-period derivative is proportional to the mass-transfer rate and the rate of angular-momentum loss \citep{Di2008}. Therefore, the negative orbital-period derivative of 4U 1820 remains mysterious for a white-dwarf binary. In this work, we propose a CB disk scenario with a cycle-mass feed to interpret the anomalous orbital-period derivative of this source. In our model, the runaway burning process during a superburst event may carry away the material of burning layer.
If a small fraction of the material feeds a CB disk around the binary rather than leaves it,
the related abrupt change in the total angular-momentum-loss rate destroy the condition for a stable mass transfer, and induce the observed negative orbital-period derivative during the recovery of the stable mass transfer.

Although the mass of the burning-carbon layer on the neutron-star surface is only $\sim5\times10^{-8}$ ${\rm M}_\odot$ during normal superburst events, we assume that some rare superbursts with a probability of 0.1\% have a burning-layer mass of $\sim5\times10^{-7}$ ${\rm M}_\odot$.
In calculation, rare superbursts are assumed to reoccurrence events with a cycle period of 10,000 yr, and 2 percent of the burning-layer mass feed into a CB disk around the binary.
In principle, photo-evaporation process by the neutron star's spin-down energy would decrease the mass of the CB disk \citep{anto14}. However, the mass-loss rate of the CB disk sensitively depends on the photo-evaporation efficiency, the binary separation, and the spin period of the pulsar \citep{alex06,owen12,ches13}. Therefore, we simply assume that the CB disk lost 0.1\% of its current mass every year. Our numerical calculation show that such a CB disk model can account for the donor-star mass, orbital period, and orbital-period derivative observed in 4U 1820.
If the rare superburst events have a recurrence time of 10,000 yr, the CB disk model predicts a 900 yr timescale in the period-decreasing stage, and gives a probability of $\sim9\%$ to observe a negative period derivative. Certainly, because of the recovery of stable mass transfer, the absolute value of orbital-period derivative should slowly decrease
which can be tested by future further observations.
In addition, because the continuum contribution of dust emission from the CB disk could observed in L band ($3-4~\rm\mu m $) \citep{spru01}. Therefore, we expect future detailed multi-waveband observations for 4U 1820 to confirm or refute our scenario.

\acknowledgments {
We are grateful to the anonymous referee for his/her constructive comments. This work was supported by the Natural Science Foundation of China under grant Nos. 11573016, 11133001 and 11333004,
the Strategic Priority Research Program of CAS under grant No. XDB09010000, the Program for Innovative Research Team (in
  Science and Technology) at the University of Henan Province, and the
  China Scholarship Council.}

\end{document}